\documentclass[10pt,preprint]{aastex}

\slugcomment{ApJL, in press}

\shorttitle{Suppression of terrestrial planet formation}
\shortauthors{Armitage}

\begin{document}

\title{A reduced efficiency of terrestrial planet formation following \\ giant planet migration}

\author{Philip J. Armitage\altaffilmark{1,2}}
\altaffiltext{1}{JILA, Campus Box 440, University of Colorado, Boulder CO 80309; pja@jilau1.colorado.edu}
\altaffiltext{2}{Department of Astrophysical and Planetary Sciences, University of Colorado, Boulder CO 80309}

\begin{abstract}
Substantial orbital migration of massive planets may occur in 
most extrasolar planetary systems. Since migration 
is likely to occur after a significant fraction of the dust 
has been locked up into planetesimals, ubiquitous migration 
could reduce the probability of forming terrestrial 
planets at radii of the order of 1~au. Using a simple 
time dependent model for the evolution of gas and solids 
in the disk, I show that replenishment of solid material 
in the inner disk, following the inward passage of a 
giant planet, is generally inefficient. Unless the 
timescale for diffusion of dust is much 
shorter than the viscous timescale, or planetesimal 
formation is surprisingly slow, the surface 
density of planetesimals at 1~au will typically be depleted 
by one to two orders of magnitude following giant planet 
migration. Conceivably, terrestrial planets may exist only in 
a modest fraction of systems where a single generation of 
massive planets formed and did not migrate significantly. 
\end{abstract}


\section{Introduction}
Radial velocity surveys \citep{mayor95,marcy00} have discovered a 
significant population of giant extrasolar planets with orbits 
interior to the snow line of protoplanetary disks 
\citep{hayashi81,bell97,sasselov00}. These discoveries -- 
together with the generally accepted theoretical belief that 
giant planet formation is difficult at such small radii 
\citep{bodenheimer00} -- suggest that massive planet formation 
is often accompanied by orbital migration, which can be driven 
by gravitational interactions with the gas disk \citep{goldreich80,lin86,lin96}. 
The $\approx 2/3$ of currently known systems that have semi-major 
axis $a < 1.5 \ {\rm au}$ are obviously prime candidates for 
having undergone migration, but theoretical models 
\citep{trilling98} suggest that even this is an underestimate. 
\cite{trilling02} have shown that the 
observed planets were probably preceded by earlier generations 
that failed to survive migration, while \cite{armitage02} estimate 
that little or no migration (defined as a fractional change in 
semi-major axis $\vert \Delta a \vert / a < 0.1$) 
occurs in as few as 15\% of systems with a surviving massive planet.

If migration is indeed ubiquitous, it is reasonable to expect 
that it will affect both the formation and the survival prospects 
of co-existing terrestrial planets. Several authors have studied 
the {\em dynamical} stability of notional terrestrial planets in 
known extrasolar planetary systems \citep{gehman96,jones01,menou02}. 
Stable orbits that could in principle harbor terrestrial planets exist 
in most of these systems. The exhaustive study by \cite{menou02} 
concludes that at least 25\% could host habitable 
terrestrial planets, while uninhabitable low mass planets 
could be present in a larger fraction of systems. 

In this paper, I investigate whether migration 
is likely to restrict the {\em formation} of terrestrial 
planets. It is fairly clear that a doomed giant planet, migrating 
inward toward eventual merger with the star, will cause any 
pre-existing low mass planets or planetesimals at smaller 
radii to be lost. What is less 
obvious is whether a subsequent generation of planetesimals (and  
eventually planets)  
could form from the potentially still massive disk present 
after the merger. To investigate this, I describe in \S2 
a simple model for the evolution of gas and solids
in the disk. In \S3, I present results for the surface 
density of solids at small radii following massive planet 
migration. I find that rather little material is likely to be 
able to repopulate the terrestrial planet forming region, and 
discuss in \S4 the implications for the frequency of terrestrial 
planets.

\section{Gas disk and dust evolution model}
The evolution of the protoplanetary disk, and the 
coagulation of dust within it to form (eventually) planets, 
remain imperfectly understood. I describe here a simple three 
component model, for the evolution of gas, dust and planetesimals, 
that is intended only to capture the broad outlines 
of the probable behavior.  

The gas component is modeled as a viscous accretion disk 
\citep{pringle81}, with disk parameters chosen to reproduce 
the typical lifetimes \citep{strom89,haisch01} and accretion rates 
\citep{gullbring98} of Classical T Tauri disks. The 
surface density $\Sigma(r,t)$ evolves according to,
\begin{equation} 
 { {\partial \Sigma} \over {\partial t} } = 
 { 3 \over r } { \partial \over {\partial r} } 
 \left[ r^{1/2} { \partial \over {\partial r} } 
 ( \nu \Sigma r^{1/2} ) \right] - \dot{\Sigma}_{\rm wind},
\label{eq_sigma} 
\end{equation}
where $\nu$ is the kinematic viscosity and the term 
$\dot{\Sigma}_{\rm wind}$ allows for mass loss in a disk wind.
For the viscosity I adopt,
\begin{equation} 
 \nu = 4.1 \times 10^{13} \ \left( {r \over {1 \ {\rm au} } } \right) \ 
 {\rm cm^2 \ s^{-1}},
\end{equation}
while for the mass loss a prescription motivated by observations 
and theoretical modeling of photoevaporation \citep{johnstone98,clarke01,matsuyama02} 
is used,
\begin{eqnarray} 
 \dot{\Sigma}_{\rm wind} & \propto & r^{-5/2}, \,\,\,\,\, r > 5 \ {\rm au} \nonumber \\
 \dot{\Sigma}_{\rm wind} & = & 0, \,\,\,\,\, r < 5 \ {\rm au}.
\end{eqnarray} 
The normalization of the mass loss is chosen such that the mass 
loss rate within 25~au is $10^{-9} \ M_\odot \ {\rm yr}^{-1}$. 
The rate of surface density loss $\dot{\Sigma}_{\rm wind}$ is 
constant in time. The initial surface density profile is,
\begin{equation} 
 \Sigma = \Sigma_0 \left( 1 - \sqrt{r_{\rm in} \over r} \right) r^{-1} e^{-r/r_0},
\end{equation}
where $r_{\rm in} = 0.067 \ {\rm au}$ is the inner disk radius and 
$r_0 = 10 \ {\rm au}$ is a scale radius chosen to truncate the initial 
profile. The constant $\Sigma_0$ is chosen such that the initial 
accretion rate at $r_{\rm in}$ is $\dot{M} = 5 \times 10^{-8} \ M_\odot \ {\rm yr}^{-1}$.
With these choices, the initial disk mass is $0.05 \ M_\odot$, and the 
disk lifetime is around 8~Myr. These values are consistent with 
observations \citep{beckwith90,strom89,haisch01}.

Solids in the disk are divided into two components, depending upon 
whether they are well-coupled to the gas (colloquially `dust') or decoupled 
(`planetesimals'). The dust component, of surface density $\Sigma_d$, 
is treated as a trace contaminant to the dominant gas component, 
with concentration $C \equiv \Sigma_d / \Sigma \ll 1$. This 
simplifies matters considerably, since the concentration obeys 
the equation \citep{clarke88},
\begin{equation} 
 { {\partial C} \over {\partial t} } = 
 { 1 \over {r \Sigma} } 
 { \partial \over {\partial r} } 
 \left( D r \Sigma { {\partial C} \over {\partial r} } \right) 
 - v_r { {\partial C} \over {\partial r} },
\label{eq_C} 
\end{equation}
where $D$ is an effective diffusion coefficient and 
$v_r$ is the radial velocity. This equation allows 
for both advection of dust with the mean disk flow, 
and diffusion in regions of the disk where there 
is a gradient in the dust concentration. Although both 
are byproducts of turbulence, there is no reason why the 
diffusion coefficient should equal the viscosity\footnote{The 
relation between $\nu$ and $D$ could be determined via 
numerical simulations, 
though the properties of disk turbulence at the radii 
of interest remain subject to uncertainty due to the low 
ionization fraction and resulting suppression of the 
magnetorotational instability \citep{gammie96,balbus00}. 
There are also constraints on the efficiency of 
mixing processes from the isotopic heterogeneity inferred 
for the Solar nebula \citep{lugmair98,boss01}.}, 
but I assume that they have the same radial scaling 
and write $D = \xi \nu$, with $\xi$ a constant.
Initially, I take $C=10^{-2}$ throughout the disk.
The disk wind is assumed to entrain dust with it, 
leaving $C$ in the material left behind unaltered.

Coagulation of the dust will produce a population 
of larger (km sized and upward) bodies which are 
decoupled from the gas and follow Keplerian orbits. 
Eventually, these may form planets (\cite{kenyon02}, 
and references therein). It is generally believed that the 
rate limiting step in this process occurs during the early 
phases of dust settling \citep{weidenschilling80,lissauer93}, and that 
rapid growth occurs through the size range where 
radial drift is rapid. In these circumstances a 
local prescription for converting dust into planetesimals 
suffices. I adopt the simple form,
\begin{eqnarray}
 { {d \Sigma_d} \over {d t} } & = & - \left( {r \over {1 \ {\rm au}} } \right)^{-3/2} 
 {\Sigma_d \over \tau_0} \nonumber \\
 { {d \Sigma_p} \over {d t} } & = & \left( {r \over {1 \ {\rm au}} } \right)^{-3/2} 
 {\Sigma_d \over \tau_0}.
 \label{eq_sigmadust}
\end{eqnarray}  
Here $\Sigma_p$ is the surface density in decoupled solid 
bodies, and $\tau_0$ is a characteristic time for dust 
settling at 1~au \citep{nakagawa86}. Observations suggest 
that significant dust evolution occurs on Myr timescales 
even at $r > 10^2 \ {\rm au}$ \citep{throop01}, which 
lends support to the theoretical conclusion that $\tau_0$ 
is short. I assume a fiducial value $\tau_0 = 10^4 \ {\rm yr}$, 
but consider the effect of varying this parameter.

If growing terrestrial planets attain masses of the 
order of $M_\earth$ prior to the dispersal of the 
gas disk, gravitational torques can once again couple 
the gas and solid components and drive rapid migration 
\citep{ward97}. By using equation (\ref{eq_sigmadust}), 
which ignores this effect, I am therefore implicitly assuming 
that the final stages of terrestrial planet formation 
occur subsequent to the loss of the disk. In the 
terminology of \cite{ward97b}, this is the `slow' 
mode of planet formation, which is often assumed to be 
appropriate for our own terrestrial planets \citep{wetherill90}. 
The possibility of an alternate, `fast' mode, in which 
Type I migration is an essential element, is briefly 
discussed in \S3.

Equations (\ref{eq_sigma}) and (\ref{eq_C}) are solved 
on a non-uniform radial grid, with 200 mesh points 
extending from 0.067~au to 
200~au. Standard numerical methods suffice for both 
equations, and can be validated using analytic 
solutions to special cases of both the surface 
density equation \citep{pringle81} and the 
contaminant equation \citep{clarke88}. Zero-torque 
boundary conditions are applied at the inner edge 
of the disk.

\section{Results}

\begin{figure}
\plotone{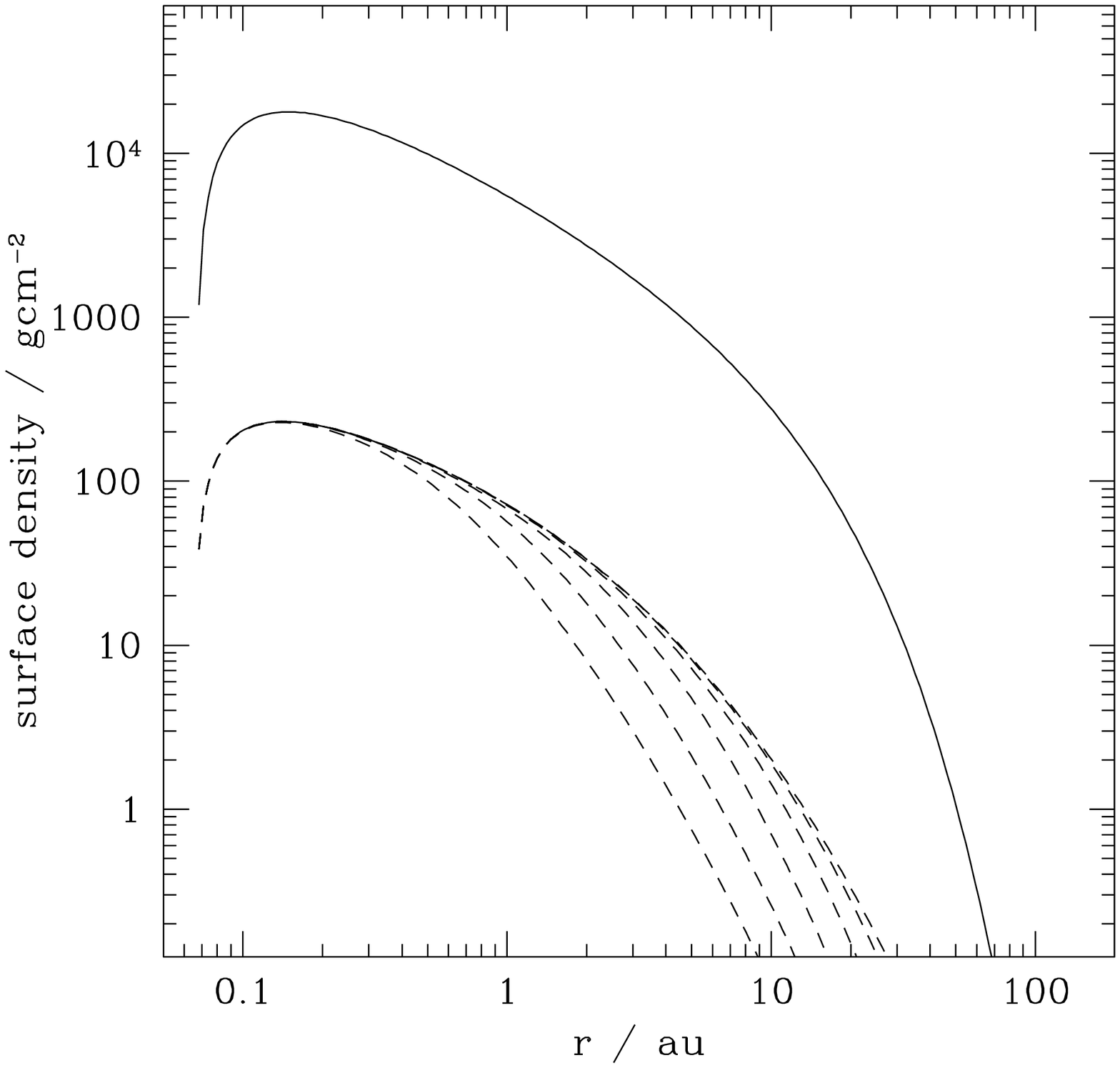}
\caption{Build up of the surface density in planetesimals for a model 
with $\tau_0 = 10^4 \ {\rm yr}$ and $\xi=1$. The dashed curves show 
results at $\log (t/{\rm yr}) = 4$, 4.5, 5, 5.5, 6 and 6.5.
The upper solid curve shows the initial gas surface density distribution.}
\label{fig1}
\end{figure}

Figure~\ref{fig1} shows the basic evolution of the 
aforementioned model. The initial disk has a gas surface 
density at 1~au that exceeds the minimum mass Solar nebula 
by a factor of about 5, and extends out to a few 10's of 
au. With dust diffusion and planetesimal formation 
parameters $\xi =1$ and $\tau_0 = 10^4 \ {\rm yr}$, 
the build up of solids decoupled from the gas is rapid 
in the inner disk, but requires of the order of a Myr 
at 5-10~au. The final radial distribution of planetesimals 
is similar but not identical to the initial gas distribution.

Once a massive planet has reached 1~au, the time remaining 
until merger is short. To a good approximation, therefore, 
the influence of giant planet migration on the distribution 
of solid material in the inner disk can be modeled by 
instantaneously sweeping the inner disk clear of 
planetesimals at some specified time $t_{\rm migrate}$. 
By varying $t_{\rm migrate}$, we can study the extent to which 
dust continues to advect and diffuse to small radii, and thereby 
form new planetesimals.
 
\begin{figure}
\plotone{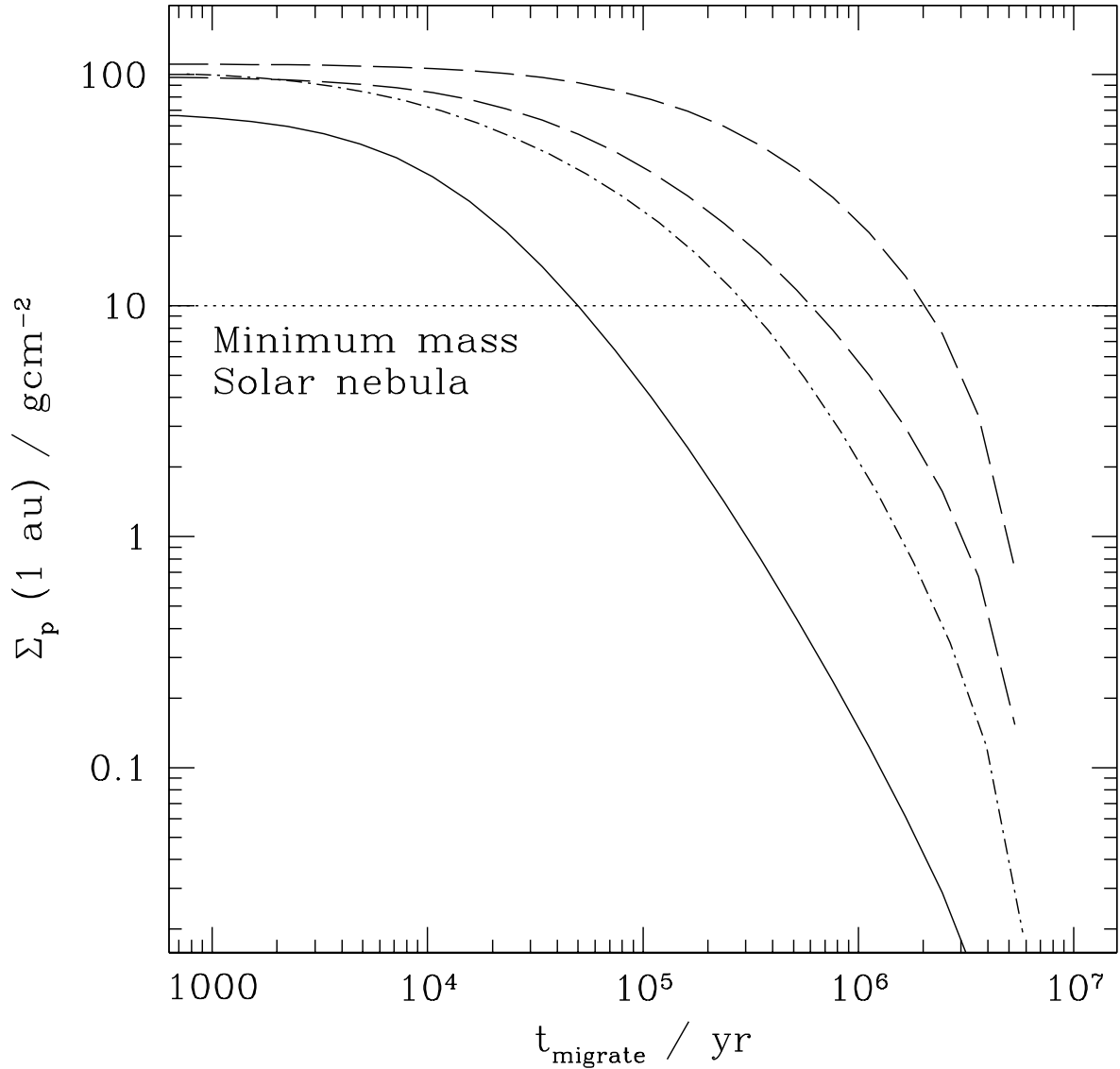}
\caption{The final surface density in planetesimals at 1~au, as a function 
of when a single massive planet migrated through the inner disk. The solid 
curve shows the results for a model with $\tau_0 = 10^4 \ {\rm yr}$ and $\xi=1$. 
The other curves each vary one parameter. The two dashed curves have longer 
planetesimal formation times ($\tau_0 = 3 \times 10^4 \ {\rm yr}$, 
$\tau_0 = 10^5 \ {\rm yr}$), and the dot-dashed curve more efficient diffusion ($\xi = 5$).
The dotted line shows an estimate of the surface density in planetesimals 
at 1~au for a minimum mass Solar nebula model.}
\label{fig2}
\end{figure}

Figure~\ref{fig2} shows $\Sigma_p$ at $r=1 \ {\rm au}$ as a 
function of $t_{\rm migrate}$. The surface density of 
decoupled solid material is evaluated at the end of the 
calculation, after the gas in the disk has been lost. 
For the fiducial parameters of $\xi =1$, $\tau_0 = 10^4 \ {\rm yr}$, 
the final surface density of planet-forming material is 
heavily suppressed unless migration occurs extremely 
early -- within the first $10^5 \ {\rm yr}$. Even 
taking into account the fact that, in the absence of 
migration, the final surface density of planetesimals 
is rather large (substantially in excess of the 
minimum mass Solar nebula value), it seems unlikely 
that a second generation of 
planetesimals would be formed in large enough numbers 
to permit terrestrial planet formation following migration.

How sensitive is this conclusion to the rather uncertain 
input parameters of the model? Although the adopted 
$\tau_0 = 10^4 \ {\rm yr}$ has some theoretical support 
\citep{lissauer93}, taking $\xi=1$ is simply a plausible 
guess. Since the concentration decreases toward small 
radii, where planetesimal formation is more rapid, more 
dust is able to repopulate the inner disk following 
migration if the efficiency of dust diffusion is 
enhanced. Increasing the timescale on which dust forms 
planetesimals obviously has the same effect. Figure~\ref{fig2} 
shows that by adopting $\xi=5$, or modestly increasing 
the planetesimal formation time to $\tau_0 = 3 \times 10^4 \ {\rm yr}$, 
the timescale on which migration could occur {\em without} 
detrimental effects on subsequent terrestrial planet 
formation can be lengthened to perhaps a Myr. Since the 
disk lifetime is almost an order of magnitude in excess 
of this \citep{strom89,haisch01}, migration occurring at a random epoch would in 
most cases still inhibit terrestrial planet formation using 
these values. This would not be true if $\tau_0$ was as 
large as $10^5 \ {\rm yr}$. However, with this $\tau_0$ a single 
e-folding time for dust destruction at 5~au is in excess 
of a Myr. Early giant planet migration would then be 
unlikely, simply because there would probably not have 
been time to form a massive planet at 5~au via core accretion 
\citep{pollack96} in the first place.

Could low mass planets form further out in the disk, and 
subsequently migrate inward via gas interactions 
to fill the void at smaller radii? This is conceivable. 
The rate of Type~I migration (in which the embedded planet 
fails to open a gap) is not limited to the 
inflow velocity of the gas, and can be extremely rapid 
for Earth mass planets at radii of the order of 1~au 
\citep{ward97}. If growing planets at somewhat larger radii 
attain such masses while the gas remains present, it is 
likely that they will suffer substantial radial 
migration, and pass through the terrestrial planet region.
What is less clear is whether migration can be halted 
to leave terrestrial planets orbiting at small radii, 
given that the gas which drives migration will 
rapidly refill the inner disk following the passage 
of a massive planet. At very small radii, interaction 
with the star or its magnetosphere \citep{lin96} provides 
a mechanism which can both halt migration, and allow further 
growth as additional low mass objects migrate inward \citep{ward97b}.
Further out, it is possible that the existence of 
sharp edges in the disk -- perhaps associated with 
the onset of efficient MHD turbulence within $r \sim 0.1 \ {\rm au}$ 
\citep{gammie96} -- could play a similar role \citep{kuchner02}. 
Additional work is needed to assess the viability of these 
possibilities, and to determine how much gas 
would be accreted during the Type~I phase. Initial 
numerical simulations by \cite{dangelo02} 
and \cite{bate02}, which do not incorporate the 
complexities in the radial disk structure discussed 
above, find that both migration and 
accretion are very rapid for masses $M \sim M_\earth$. 
This would suggest that Earth mass planets migrating 
through the disk would accrete significant gaseous 
envelopes. Simulations which include a treatment 
of the energy equation are needed to determine 
whether this is a robust result.

Migration due to interactions with the gas disk is almost 
inevitable unless planets happen to form very close to the 
moment when the gas is dispersed. It is possible that 
further orbital evolution occurs subsequently, driven 
by purely gravitational interactions with planetesimals \citep{murray98}
or other massive planets \citep{lin97,ford01,marzari02}. These  
processes could also impact the probability of terrestrial planet 
formation. In particular, if dynamical interactions are 
the cause of extrasolar planet eccentricities, then an unstable 
multiple planet system must be a common outcome of the 
planet formation process. The relevant question for terrestrial 
planet formation is then not which orbits are unstable 
in the present-day system, but which orbits were swept 
free of planetesimals during the early chaotic evolution 
of a massive planet system. This could be substantially 
more restrictive of the possibilities for habitable terrestrial 
planets than the constraints derived so far.

\section{Discussion}
The main technical result of this paper is that inward 
migration of massive planets, if it occurs more than $\sim 10^6 \ {\rm yr}$ 
after the formation of the protoplanetary disk, is likely to 
suppress the subsequent formation of terrestrial planets at 
radii of the order of 1~au. This suppression occurs because 
the gas that flows in to the inner disk, after the migrating 
planet has merged with the star, is depleted of dust as a 
result of already having formed planetesimals at larger 
radii. It is possible to choose parameters such that this 
conclusion can be evaded, but this requires either slow 
conversion of dust to planetesimals (in which case it is 
hard to see how the migrating massive planets could have formed, 
at least via core accretion), or very rapid diffusion of dust on a 
timescale an order of magnitude shorter than the disk's viscous timescale. 

For future targeted searches of nearby stars for low mass planets, the 
direct implication is that stars with `hot Jupiters' -- massive 
planets at extremely small orbital radii of $a \sim 0.1 \ {\rm au}$ 
or less -- may have been unable to form terrestrial planets, even 
though there are dynamically stable orbits for habitable 
planets in those systems \citep{jones01,menou02}. Whether there 
are broader implications for the likely frequency of terrestrial 
planets depends first upon how common migration is, and second on 
whether there are stars that form terrestrial planets without 
forming any giant planets. The latter class of systems could 
certainly be common, though as their numbers are entirely 
unconstrained by existing observations any estimate is 
necessarily speculative.

How common migration is among systems that form massive planets 
depends primarily on how fast migration occurs. If 
inward migration from a few au to merger with the star is rapid 
(compared to the disk lifetime), then it requires fortuitous 
timing to form a giant planet at a moment, close to the disk 
dispersal time, when it will not migrate significantly. Numerical 
simulations \citep{bate02} indicate that the migration timescale 
(at least in the simplest disk models) is short for planets formed 
close to the snow line, and this is reflected in migration being the 
rule rather than the exception in statistical models of the extrasolar 
planet distribution \citep{armitage02,trilling02}. Among stars 
that either now possess massive planets, or {\em once formed} 
massive planets that were subsequently lost, migration 
through to merger is likely to have occurred in the 
majority of systems \citep{trilling02}. As a result, it is 
possible that a high efficiency of massive planet formation 
could depress the frequency of terrestrial planets to a 
level substantially below the massive planet frequency.

\acknowledgements

I thank Nick Schneider for posing difficult questions that prompted 
this work, and the referee for extremely helpful suggestions.

\newpage

\end{document}